\documentstyle[12pt,epsfig]{article}
\begin{document}

%     global definitions go here

\def\etal{{\em et al.}}
\def\nue{\nu_{e}}
\def\nuebar{\overline{\nu}_{e}}
\def\numu{\nu_{\mu}}
\def\numubar{\overline{\nu}_{\mu}}
\def\nutau{\nu_{\tau}}
\def\nutaubar{\overline{\nu}_{\tau}}
\def\nux{\nu_x}
\def\nuxbar{\overline{\nu}_{x}}

\def\dmsq{\Delta m^{2}}
\def\sinsq{{\rm sin}^2 2\theta}
\def\isotope#1{\mbox{${}^{#1}$}}                  % \isotope{52}Fe
\def\ra{\rightarrow}
\def\lra{\leftrightarrow}
\def\units#1{\hbox{$\,{\rm #1}$}}                % 3.2\units{\mu s}

%Other definitions used in  the Chooz paper but not here
%\def\Reines{$\overline{\nu}_e\,+\,p\,\rightarrow\,e^+\,+\,n$}
%\def\SN#1E#2 {\mbox{$#1\times10^{#2}$}}      % sci.not. \SN3.2E-
%4[blank]
%\def\SNE#1+#2E#3 {\mbox{$(#1\pm#2)\times10^{#3}$}} % sci.not w/errors 
%\SN1.24+0.34E-2[blank]
%

%\begin{titlepage}

%\centerline{ \bf  DRAFT  June 3rd 1999}

\vspace{1truecm}
\begin{center}

{\bf Determination of neutrino incoming direction \\
in the CHOOZ experiment and its application to  \\
Supernova explosion location by scintillator detectors} \\

%********************************** other titles (commented)
%{\bf Location of source in SN neutrino detection \\ by scintillator 
%experiments.}
%\vspace{1truecm}
%\\
%*********************************
%{\bf Determination of neutrino incoming direction \\
%in the CHOOZ scintillator detector and \\
%SN explosion location.}
%\vspace{1truecm}
%\\  
%*********************************

\end{center}

\begin{center}
M.~Apollonio$^c$,
A.~Baldini$^b$,
C.~Bemporad$^b$,
E.~Caffau$^c$,
F.~Cei$^b$,
Y.~D\'eclais$^{e,1}$,
H.~de~Kerret$^f$,
B.~Dieterle$^i$,
A.~Etenko$^d$,
L.~Foresti$^b$,
J.~George$^i$,
G.~Giannini$^c$,
M.~Grassi$^b$,
Y.~Kozlov$^d$,
W.~Kropp$^g$,
D.~Kryn$^f$,
M.~Laiman$^e$,
C.E.~Lane$^a$,
B.~Lefi\`evre$^f$,
I.~Machulin$^d$,
A.~Martemyanov$^d$,
V.~Martemyanov$^d$,
L.~Mikaelyan$^d$,
D.~Nicol\`o$^b$,
M.~Obolensky$^f$,
R.~Pazzi$^b$,
G.~Pieri$^b$,
L.~Price$^g$,
S.~Riley$^{g,h}$,
R.~Reeder$^h$,
A.~Sabelnikov$^d$,
G.~Santin$^c$,
M.~Skorokhvatov$^d$,
H.~Sobel$^g$,
J.~Steele$^a$,
R.~Steinberg$^a$,
S.~Sukhotin$^d$,
S.~Tomshaw$^a$,
D.~Veron$^f$,
and V.~Vyrodov$^f$
\end{center}

\begin{center}
$^a${\em   Drexel University                    }  \\
$^b${\em   INFN and University of Pisa          }  \\
$^c${\em   INFN and University of Trieste       }  \\
$^d${\em   Kurchatov Institute                  }  \\
$^e${\em   LAPP-IN2P3-CNRS Annecy               }  \\
$^f${\em   PCC-IN2P3-CNRS Coll\`ege de France   }  \\
$^g${\em   University of California, Irvine     }  \\
$^h${\em   University of New Mexico, Albuquerque}  \\
$^1${\em   Now at IPN-IN2P3-CNRS, Lyon}  \\
\end{center}
%\centerline{\large \sl PACS: \large 14.16.P, 28.50.Hw}

\begin{abstract}

\noindent
The CHOOZ experiment
\footnote{The CHOOZ experiment is named after the new nuclear power 
station operated by \`Electricit\`e de France (EdF) near the village of 
Chooz in the Ardennes region of France.}
has measured the antineutrino flux at  about 
1\units{Km} from two nuclear reactors  to search for possible 
$\nuebar \ra \nuxbar$ oscillations with mass-squared  differences as low 
as $10^{-3}\units{eV^2}$ for full mixing. We show that the analysis 
of the $\sim 2700  \ \nuebar$--events, collected by our liquid 
scintillation detector, locates the antineutrino source 
within a cone of half-aperture $\approx 18^{\circ}$ at the 
$68\% \ {\rm C.L.}$ . We discuss the implications of 
this  result for locating a supernova explosion.
\end{abstract}

%{\em Key words:} reactor, neutrino mass, neutrino mixing, neutrino 
%oscillations,supernova,supernova direction.

\section{Introduction}

Locating a $\nu$--source in the sky is of primary
importance in the case of galactic supernova (SN) explosions; 
particularly if the SN is not optically visible, either because it 
is hidden behind the dust of the galactic disk, or because the 
light emission follows the neutrino burst by hours or days. In this latter 
case, an early SN detection and location by neutrinos could allow  
observation of the evolution of the  first optical  stages.

Several pointing methods have been extensively discussed and 
compared in the literature \cite{Burrows},\cite{Vogel1}:

\begin{itemize}
\item  the identification of $\nue + {\rm e}$ scattering events 
(a minority of the total number of events in most detectors);

\item the pronounced anisotropy of the charged-current (CC) $\nue + d$ and 
$\nuebar + {\rm d}$ reactions; 

\item the slight positron anisotropy in the CC reaction
$\nuebar+ {\rm p} \ra {\rm n} + {\rm e}^{+}$; 

\item the triangulation between at least three detectors. 

\end{itemize}
$\nue + {\rm e}$ scattering events have an anisotropic angular
distribution and, when observed in large detectors with good angular 
information, like the water \v Cerenkov detectors,  permit localization of  
the SN. This technique was successfully applied to solar neutrinos 
 by the KAMIOKANDE and the SUPERKAMIOKANDE (SK) experiments\cite{KSK}.
It has been shown\cite{Vogel1} that this method can determine
the direction of a Supernova, at a distance of 10 Kpc, within a
 $\sim 5^\circ$ cone 
(one--sigma angular width).
The other methods mentioned above have less precise pointing capabilities.

We  show in this paper that, in the case of
large scintillation detectors, another tool, based on the 
inverse-beta-decay reaction
\begin{eqnarray}
\label{reaction}
\nuebar + \rm{p} \ra \rm{n} + \rm{e}^{+},
\end{eqnarray}

\noindent
is also capable of providing a good determination of Supernova direction.
Although the positron is emitted almost
isotropically  in this reaction, 
the neutron has an energy-dependent maximum angle of emission 
and an associated Jacobian peak in its angular distribution. The neutron 
therefore retains a memory of the neutrino direction, which survives 
even after collisions with the protons of the 
moderating scintillator medium.

A first study along this line 
\cite{Bemporad} was performed for the MACRO experiment and its SN 
detection capabilities, with results which were not encouraging. 
However, the 
single-vessel structure of the CHOOZ experiment and its superior energy 
and position resolutions appeared more promising.  
An average displacement of neutrons with respect to positrons
  has been observed 
in previous reactor neutrino experiments \cite{Gosgen,Bugey}.
In CHOOZ we were able 
to test this neutron recoil 
method and to use it for locating the reactor-neutrino 
source.

The CHOOZ experiment has  also been a  good test bench for our 
Montecarlo simulations of the slowing down and capture of neutrons in 
a scintillator, and for
our  methods of event reconstruction in a single-vessel 
structure observed by photomultipliers. We believe that extrapolation 
from CHOOZ to much larger volume scintillator detectors can be reliably 
made, and we therefore evaluate the SN locating 
capabilities of future scintillator detectors in the $1000\units{tons}$ 
range, like BOREXINO and KAMLAND.

\section{Antineutrino detection in scintillator experiments}
\label{Antidet}

$\nuebar$'s are detected in scintillator experiments via
 inverse-beta-decay (Reaction \ref{reaction}).
The $\nuebar$ signature is a delayed coincidence between the prompt
$e^+$ signal (boosted by the two 511\units{KeV} annihilation gamma 
rays)
and the signal from the neutron capture. The target material is a
hydrogen--rich (free protons) paraffinic liquid scintillator; in CHOOZ 
the scintillator is loaded with gadolinium. 

\subsection{Positron displacement} 

The cross section for the $\nuebar$ capture process 
\cite{Xsection,Vogel1,Vogel2} can be written 
for low energy antineutrinos as
\begin{eqnarray}
\frac{d\sigma}{d\Omega}=\frac{G^2_f}{4 \pi^{2}} p_{e^+} E_{e^+}
\{ (1 + 3 F_4^2) + (1 - F_4^2)cos \theta_{e^+} \}
\end{eqnarray}

With the value of $F_4$ obtained from $\beta$-decay one derives the 
positron angular distribution in the laboratory 
\begin{eqnarray}
\label{angpos}
\rm{P}(\theta_{e^+}) = \rm{Constant} \times (1-0.102 \, cos\theta_{e^+}) 
\end{eqnarray}

This tiny anisotropy 
could in principle be employed if one had an extremely large
number of events. But this is not normally the case for SN explosions nor 
was it for the CHOOZ long-baseline reactor-neutrino experiment.
In the case of CHOOZ, by averaging Eqn. \ref{angpos} 
over the solid angle and
positron spectrum, one obtains an 
average positron displacement with respect to the interaction
of $\sim -0.05\units{cm}$ which, as we will see later,
is not measurable with our spatial resolution
and statistics.
In \cite{Vogel1,Vogel2} the dependence of the $cos\theta_{e^+}$ coefficient
on the positron energy is discussed. Such a dependence does not
affect the conclusions of this paper because the average positron 
displacement remains negligably small.

\subsection{Neutron displacement}
\begin{figure}[htb]
  \begin{center}
    \mbox{\hspace{-1.0cm}\epsfig{file=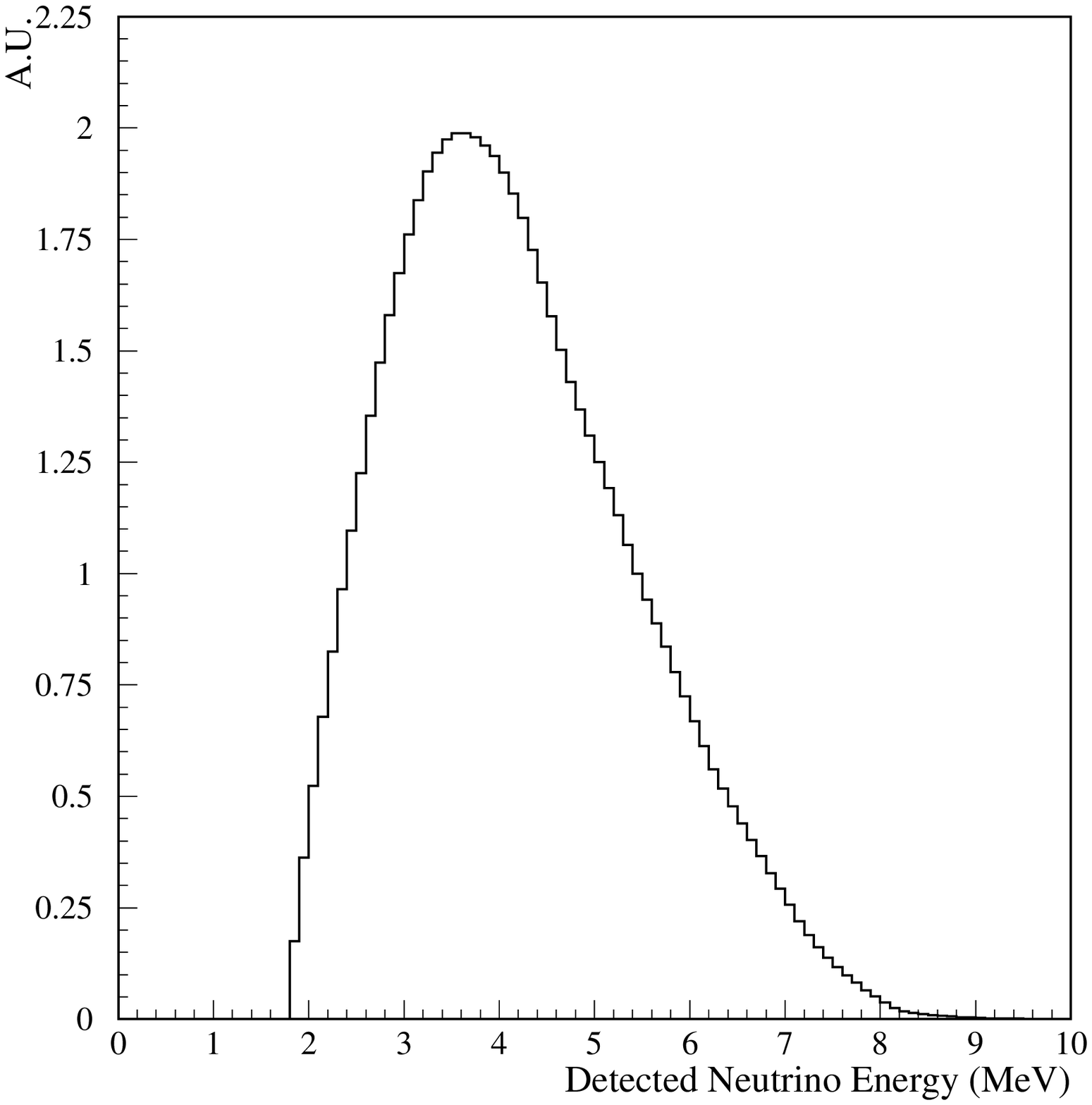,width=0.6\linewidth}
          \hspace{-0.5cm}\epsfig{file=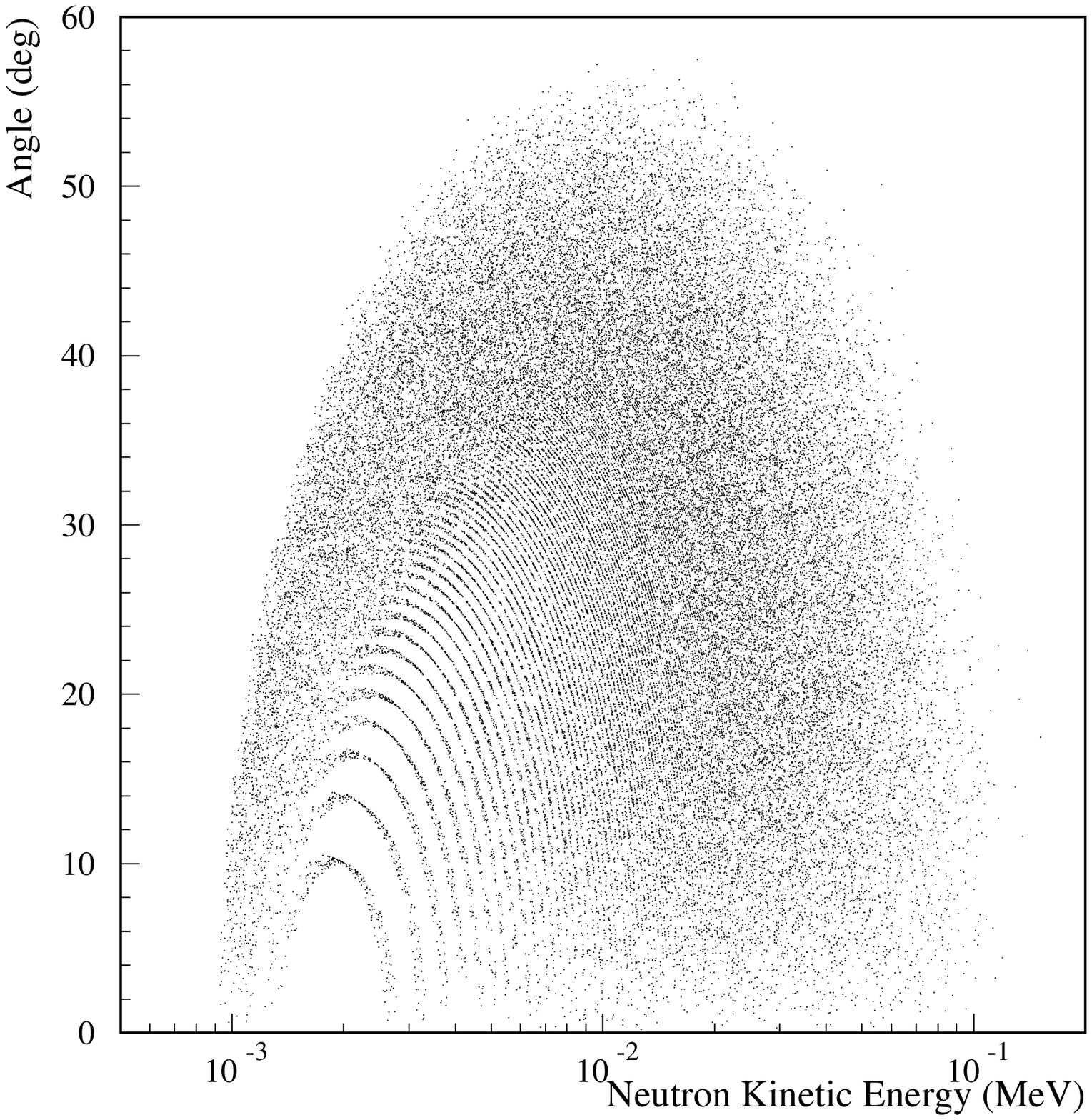,width=0.6\linewidth}}
    \caption{\small Detected reactor neutrino energy spectrum
             (left) and neutron angle with respect to the incident
             neutrino direction versus its kinetic energy(right).}
    \label{tvsn}
  \end{center}
\end{figure}

%\begin{figure}[htb]
%\begin{center}
%\mbox{\epsfig{file=tn.ps,height=4.0in,%
%bbllx=30bp,bblly=399bp,bburx=534bp,bbury=665bp}}
%\vspace{-0.2truecm}
%\caption{Neutron angle with respect to the incident neutrino
%direction versus its kinetic energy for a reactor neutrino
%energy spectrum.}
%\label{tvsn}
%\end{center}
%\end{figure} 
Figure \ref{tvsn} (right) shows the scatter plot $\theta_n$ (angle with
respect to the incident neutrino direction) versus 
$\rm{T_n}$ (kinetic energy) for neutrons emitted in Reaction
\ref{reaction} with our detected reactor-neutrino spectrum (left).
The neutron kinetic energy extends up
to $\sim 100$ KeV and the angle with respect to
the incoming neutrino direction is limited to values below $\sim 55^\circ$.
The separated curves visible in the lower-left part of the picture are caused
by the $\nuebar$ energy binning and by the logarithmic scale
adopted for the x-axis. They show how the $\theta_n - \rm{T_n}$ 
dependence changes as a function of the $\nuebar$ energy.

The neutron slowing-down phase\cite{Persico}, in which its energy
is reduced to the thermal equilibrium value,
is a non-isotropic process 
which preserves a memory of the initial
neutron direction. In each elastic scattering the average cosine of the outgoing neutron
with respect to the incoming direction is:

\begin{eqnarray}
\label{cosine}
\overline{\rm{cos}\theta_{\rm n}} = \frac{2}{3{\rm A}}
\end{eqnarray}

\noindent
where A is the atomic number of the scattering nucleus.
 The direction is therefore best preserved
by scatterings on hydrogen, which in fact is
 the most probable occurrence since the
elastic scattering cross section on hydrogen is larger than that on carbon
for energies below $1\units{MeV}$.

Slowing-down is an extremely efficient process in which 
the neutron energy rapidly decreases (an average factor of 
two for each scattering
on hydrogen). Since the scattering cross section is a decreasing function 
of the neutron energy, it follows that the neutron mean free path
($\lambda_{\rm s}$) rapidly diminishes during moderation.
As a matter of fact
the distance from the production point travelled by neutrons before
being thermalized is determined by the first two or three scatterings, 
in which a memory of the initial direction is preserved,
according to Eqn.~\ref{cosine}.

The subsequent (isotropic) diffusion process does not alter the average 
neutron displacement along the neutrino direction. In  CHOOZ 
this average displacement is $\sim 1.7\units{cm}$.

In Table~\ref{neutrons} we give some of the parameters characterizing
the moderation and diffusion phases ($\hat{n}_s$: average number 
of scatterings; $t_d$:
time duration; $\sqrt{\overline{r^2}}$: 
average square distance, which does not include
the neutron angular distribution of Reaction \ref{reaction}) 
computed for neutrons
from reactor-neutrino interactions in the CHOOZ 
liquid scintillator.

\begin{table}[htb]
    \caption{\small Parameters characterizing the 
    neutron moderation and diffusion phases in CHOOZ (see text)}
    \label{neutrons}
  \begin{center}
    \leavevmode
    \begin{tabular}{|c|c|c|c|}

      \hline
       & $\hat{n}_s$ & $t_d$ & $\sqrt{\overline{r^2}}$ \\
      \hline
       Moderation & $10 - 14$ & $9\units{\mu s}$ & $\approx 6 {\rm cm}$\\ 
      \hline
       Diffusion  & $\approx 20$ & $\sim 30\units{\mu s}$  & 
       $\approx 3\units{cm}$  \\ 
      \hline

    \end{tabular}
  \end{center}
\end{table}

\section{Description of the CHOOZ Experiment}

A description of the CHOOZ experiment,  its analysis methods, and 
 its initial results has been previously  published \cite{Chooz}. 
Here we recall only the  
points needed  for the present discussion.

\subsection{The neutrino source}

The Chooz power station has two pressurized-water reactors with a 
total thermal power of $8.5\units{GW_{th}}$.  The average direction
of the two reactors in the CHOOZ polar coordinate system was measured
by standard surveying techniques
to be $\phi = (-50.3\pm0.5)^\circ$ and $\theta = (91.5\pm0.5)^\circ$, where
$\theta$ is the zenith angle and the origin of the azimuthal ($\phi$)
coordinate is arbitrarily fixed.

\subsection{The Detector}
\label{Detector}

The Gd-loaded scintillator target is contained in an acrylic vessel of 
precisely
known volume immersed in a low-energy calorimeter made of unloaded
liquid scintillator. Gd was chosen becasue of its large neutron-capture
cross section and large neutron-capture $\gamma$-ray energy release
($\sim8\units{MeV}$, well separated from the natural radioactivity
background).

The detector is made of three concentric regions:
\begin{itemize}
\vspace{-1mm}
\item a central 5--ton target in a transparent acrylic container
filled with a 0.09\% Gd--loaded scintillator (``region 1'');
\vspace{-1mm}
\item an intermediate 17--ton region (70\units{cm} thick) equipped with 
192
eight--inch PMT's (15\% surface coverage, $\sim 130$ photoelectrons/MeV) 
used to protect the target from PMT radioactivity and 
to contain the gamma rays from neutron capture (``region 2'');
\vspace{-1mm}
\item an outer 90--ton optically-separated active cosmic--ray muon veto
shield (80\units{cm} thick) equipped with two rings of 24 eight--inch 
PMT's
 (``region 3'').
\end{itemize}

The detector is simple and easily calibrated, and its behaviour can be
well checked. Six laser flashers are installed in the three regions
together with calibration pipes to allow the introduction of radioactive
sources. 

Particularly important for this paper are the neutron
calibrations performed by using $^{\rm 252}{\rm Cf}$, 
a spontaneous fission source
emitting several prompt, energetic neutrons ($\overline{E} \sim 2\units{MeV}$)  
and $\gamma$ rays with energies up
to $\sim 10\units{MeV}$.

The detector can be reliably simulated by the Montecarlo
method .

\section{Data analysis}

\subsection{Event reconstruction}
\label{Erec}

The determination of the  direction to the reactors 
relies on the energy and position reconstruction of individual events.

The method presented here is based on the
data obtained from the VME ADC's 
(192 PMT's  divided into 24 groups (``patches'') of 8 PMT's each)
viewing regions 1 and 2.
We checked that this grouping does
not significantly affect the energy and position determination.
Using the additional information provided
by the TDC's might improve the position resolution. 

Each event is reconstructed by minimizing the  
Poissonian $\chi^2$
\begin{eqnarray}
\label{chi2}
\chi^2 = 2 \times \sum_{i=1}^{24} [ (N^{i}_{th} - N^{i}_{obs}) + 
N^{i}_{obs} \times log(\frac{N^{i}_{obs}}{N^{i}_{th}}) ]
\end{eqnarray}
where $N^{i}_{obs}$ and $N^{i}_{th}$ are the measured and expected
numbers of photoelectrons of the ${\rm i}_{th}$ patch,
for a given event's energy and position.

A Poissonian $\chi^2$ is used due to the frequent occurrence
of  low numbers of photoelectrons in some PMT patches; this
demands the application of the correct statistics.

The conversion factors from the measured ADC charges to the numbers
of photoelectrons are obtained for each patch from the single-%
photoelectron peak position measured by flashing the central
laser  at very low intensity.

The patch electronic gains have
differences caused by the behaviour of the active electronic components
(fan-in, fan-out etc. ) present after the PMT's. The
patch's relative gains are measured frequently using 
the 8 MeV absorption peak of the neutrons emitted by
a $^{\rm 252}{\rm Cf}$ source placed at the center of the detector.
Corrections are
applied for the different solid angles
of the various patches.

The predicted number of photoelectrons for each patch is computed 
by considering a local deposit of energy, resulting in a number
of visible photons that are tracked to the PMT's 
taking into account the different attenuation lengths 
of region 1 and 2 scintillators.

To reduce computing time PMT's are considered to be flat and
Rayleigh scattering
is neglected. Despite these simplifications the time
needed for reconstructing one event is of the order of one
second on a Pentium processor.

%The time evolution of the attenuation length of the region
%1 scintillator is followed by using the empirical formula:
%
%\begin{eqnarray}
%\label{atten}
%\lambda(t) = \frac{\lambda(0)}{1 + \alpha t} 
%\end{eqnarray}
%
%where $\lambda(0) = 512$ cm and $\alpha = \frac{1}{294}~{\rm d}^{-1}$
%are obtained by a fit to the attenuation length
%values (see fig. \ref{attenuation} ) directly measured in the apparatus
%by moving a $^{\rm 252}{\rm Cf}$ source along the central
%calibration pipe.
%Formula \ref{atten} implies an exponential decrease with time 
%for the total light emitted 
%by an ionizing source at the center of the detector. 

\subsection{Montecarlo simulation}
\label{Msim}

The detector response was simulated by means of the CERN {\sc geant} 
package. The complete geometry was taken into account. 
Particular attention was devoted 
to low-energy neutrons (${\rm E_n} < 10$ MeV) by using
specially designed routines which modeled
the  elastic scattering and capture
on individual elements of the scintillators
and acrylic vessel,
employing
the relevant experimental cross sections \cite{Relcross}.
Neutron capture on Gadolinium was simulated for the two
main isotopes: $^{155}{\rm Gd}$ and $^{157}{\rm Gd}$.
Following a capture on Gd either two monoenergetic photons
 or a varying number
of photons, obtained by considering transitions to intermediate
energy levels with a probability proportional to the cube
of the energy difference, were emitted \cite{Capture}.  

Collection of the light generated by local energy deposits was performed 
under the same approximations described in section \ref{Erec}. 

\begin{figure}[htb]
\begin{center}
\mbox{\epsfig{file=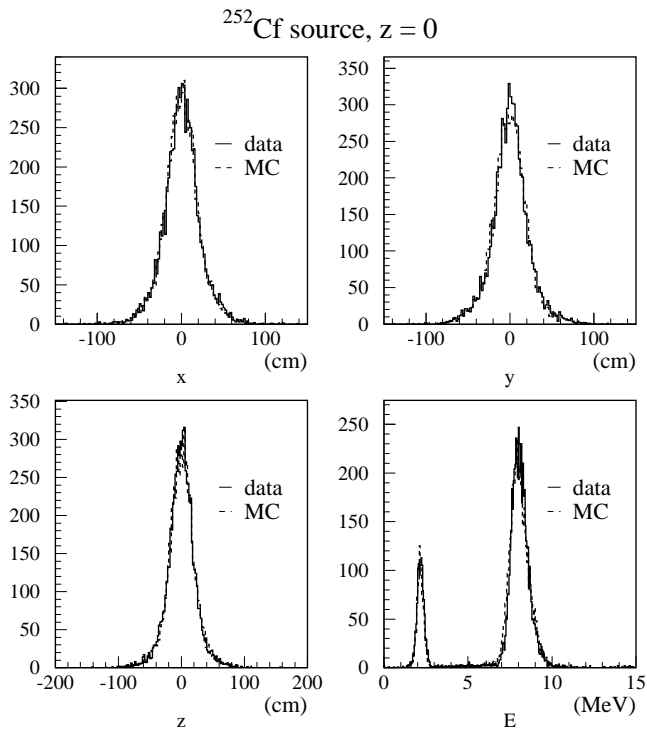,height=4.0in}}
\vspace{-0.2truecm}
\caption{Reconstructed position and energy for neutrons emitted
by a californium source at the center of the apparatus.}
\label{Cf}
\end{center}
\end{figure} 

In Figure \ref{Cf} we show the reconstructed position and energy 
for neutrons emitted by a $^{252}{\rm Cf}$ source at the apparatus
center (continuous line). The agreement with the superimposed Montecarlo
simulation (dashed line) is very good.

\subsection{Candidate selection and comparison with the Montecarlo
simulation}

The candidate selection is based on the following
requirements:

\begin{itemize}

\item energy cuts on the neutron candidate (6 -- 12 MeV) and 
on the ${\rm e}^+$ (from hardware threshold energy ${\rm E}_{thr}
\sim 1.3~{\rm MeV}$ to  8 MeV ),

\item a time cut on the delay between the ${\rm e}^+$ and the neutron
(2 -- 100 $\mu{\rm s}$),

\item spatial cuts on the ${\rm e}^+$ and the neutron (distance from
the PMT wall $> 30$ cm and distance between the  n and 
 ${\rm e}^+ \rm{vertices} < 100$ cm).

\end{itemize}

The total number of candidates for the complete data taking period 
with reactors on is $\sim2700$. We also 
collected $\sim200$ background events during two reactor-off
periods: in the middle (October 1997) and right after the end (February 1998)
of the reactor-on data taking period.
 
\begin{figure}[htb]
\begin{center}
\mbox{\epsfig{file=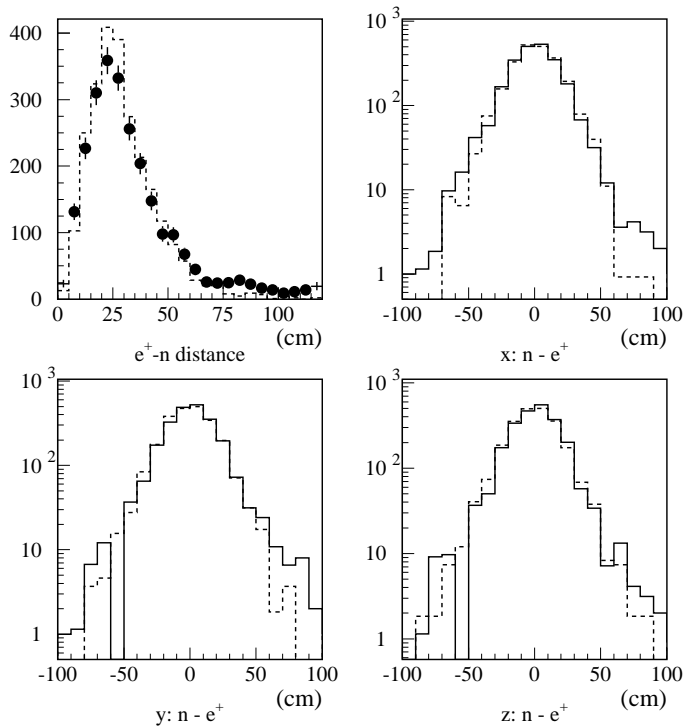,height=4.0in,%
bbllx=60bp,bblly=140bp,bburx=540bp,bbury=670bp}}
\vspace{-0.2truecm}
\caption{Background-subtracted distributions of
the difference between the coordinates of neutrons and positrons,
and the positron-neutron distance for the candidates (continuous lines)
and the expected Montecarlo distributions (dashed lines).}
\label{Cdiff}
\end{center}
\end{figure} 

Figure \ref{Cdiff} shows the background-subtracted distributions of
the difference between the coordinates of neutrons and positrons
for the candidates (continuous lines). Also shown is the 
positron-neutron distance distribution.
The expected Montecarlo distributions (dashed lines) are superimposed
on the experimental data.
The discrepancies are  
attributed to the simplified light collection scheme adopted in the
Montecarlo. 
%We also point-out that, by utilizing the difference
%of the neutrons and positrons positions, possible effects in the events
%reconstruction caused by calibration constants obtained by using sources
%not exactly positioned at the apparatus center cancel out. 
We also point out that, by utilizing the difference between the 
neutron and positron positions, most systematic errors in the 
absolute reconstructed coordinates are canceled out.

For this analysis we required the combined thermal power of both
reactors to be greater than 3 GW in order to increase the signal
to background ratio. The candidates reduce to $\sim 2500$ after this
condition is applied.

\section{Location of the reactors by neutrino events}

\subsection{The technique used}
\label{technique}

As seen in section \ref{Antidet}, while the average positron displacement
with respect to the neutrino interaction point is not measurable, 
a sizable average displacement of the neutron capture point along
the neutrino incoming direction is predicted. 
In order to measure the direction
of this average displacement we define for each neutrino candidate 
the unit vector $\hat{\rm X}_{pn}$ 
having its origin at the positron 
reconstructed position and pointing to the neutron capture location.

\begin{figure}[htb]
\begin{center}
\mbox{\epsfig{file=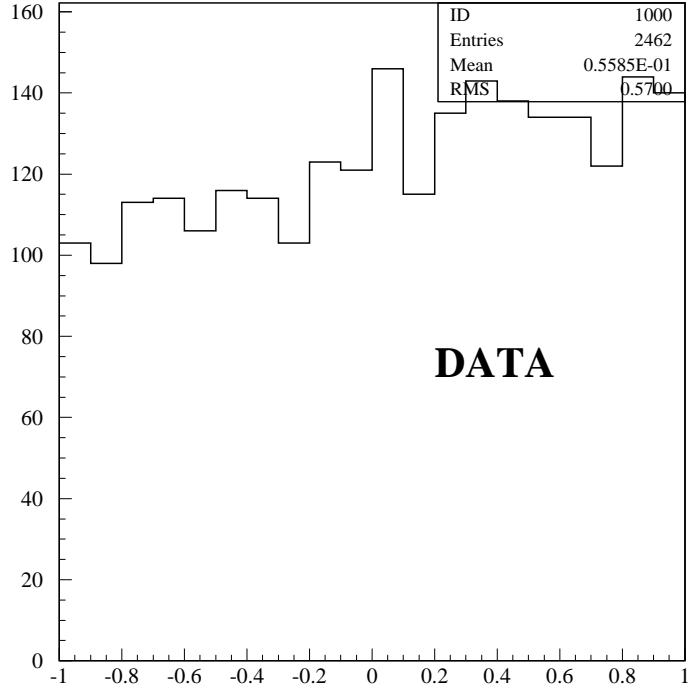,height=4.0in,%
bbllx=60bp,bblly=139bp,bburx=530bp,bbury=655bp}}
\vspace{-0.2truecm}
\caption{Distribution of the projection of the positron-neutron
unit vector along the known neutrino direction.}
\label{Coscan}
\end{center}
\end{figure} 

\begin{figure}[htb]
\begin{center}
\mbox{\epsfig{file=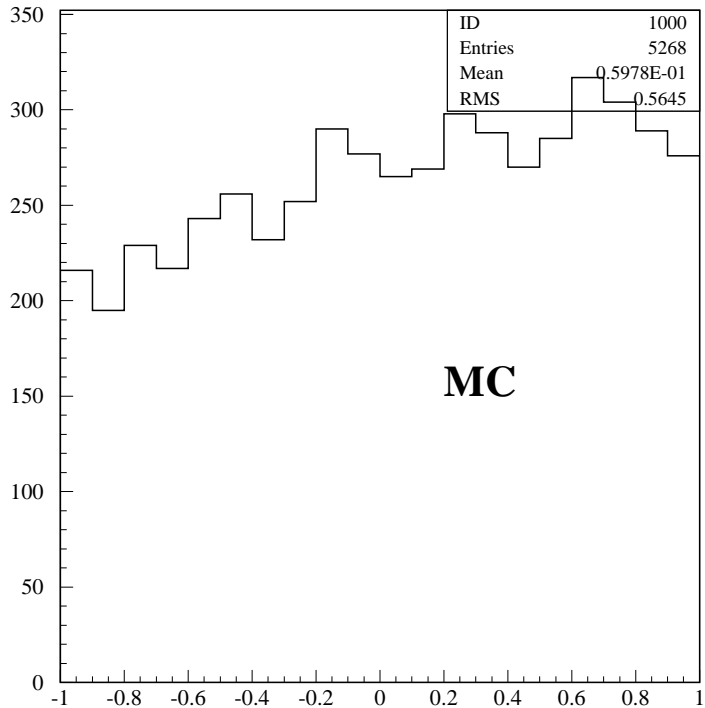,height=4.0in,%
bbllx=50bp,bblly=150bp,bburx=530bp,bbury=655bp}}
\vspace{-0.2truecm}
\caption{Distribution of the projection of the Montecarlo positron-neutron
unit vector along the known neutrino direction.}
\label{Cosmc}
\end{center}
\end{figure}

The distribution of the projection of this vector along
the known neutrino direction 
is shown for candidates in figure \ref{Coscan}.
This distribution can be compared with  Fig.~\ref{Cosmc}
which shows the distribution  expected from a Montecarlo
simulation with higher statistics
($\sim$5,000 events).

Both distributions are forward peaked athough 
their R.M.S. values (0.570 and 0.565, respectively) 
 are not far from that of a flat distribution 
($\sigma_{flat} = 1/\sqrt{3} \simeq 0.577$).

We define $\vec{\rm p}$ as the average of vectors
$\hat{\rm X}_{pn}$
\begin{eqnarray}
\vec{\rm p} = \frac{1}{N} \sum_i {\hat{\rm X}}_{pn}
\end{eqnarray} 
The measured neutrino direction is the direction of $\vec{\rm p}$.

In order to evaluate the uncertainty in this value
we need to know  the expected distribution of $\vec{\rm p}$.
Let us assume the neutrino direction lies along the z axis.
$\vec{\rm p}$ is the sum (divided by N) of N variables of which we know the
average ( $\vec{\rm p}$ itself $= (0,0,\mid \vec{\rm p} \mid$)and $\sigma$
(we can safely assume $\sigma = 1/\sqrt{3}$ for the three 
components, as seen above).
From the central limit theorem it follows therefore that the distributions
of the three $\vec{\rm p}$ components are gaussians with
$\sigma = 1/\sqrt{3N}$ centered at $= (0,0,\mid \vec{\rm p} \mid)$.
An uncertainty on the measurement of the neutrino direction can therefore
be given as the cone around $\vec{\rm p}$ which contains $68\%$
of the integral of this distribution. 

Another direction estimator we tried to use is the simple average
of the differences between the neutron and positron reconstructed
positions:
\begin{eqnarray}
\vec{\rm q} = \frac{1}{N} \sum_i ({\vec{\rm X}}_n - {\vec{\rm X}}_p)
\end{eqnarray} 
The results obtained  using this estimator are very close to those
one gets with the first one, although the corresponding uncertainties are
always a bit ($\sim 10 \%$) larger (this is also true when the methods
are applied to Montecarlo generated events). 
In the following we will therefore quote only the results obtained
with the ${\vec{\rm p}}$ estimator.
 
\subsection{Experimental result}

Table \ref{Results} shows our result
for the measured direction, compared with the Montecarlo predictions for a 
sample with the same statistics. The measured direction is in agreement
(with a $16 \%$ probability)
with that expected while it has
a negligible probability of being a fluctuation of an 
isotropic distribution. 

The measured average positron--neutron displacement for
candidates turns out to
be $1.9 \pm 0.4 \rm{cm}$, in agreement with the predicted value.

\begin{table}[htb]
    \caption{\small Measurement of neutrino direction: data and Montecarlo}
    \label{Results}
  \begin{center}
    \leavevmode
    \begin{tabular}{|c|c|c|c|c|}

      \hline
       & $\mid \vec{\rm p} \mid$ & $\phi$ & $\theta$ & Uncertainty \\
      \hline
       Data & 0.055 & $-70^\circ$ & $103^\circ$ & $18^\circ$ \\ 
      \hline
       MC   & 0.052 & $-56^\circ$ & $100^\circ$ & $19^\circ$ \\ 
      \hline

    \end{tabular}
  \end{center}
\end{table}
%[fissione spontanea, emissione di gamma
%prompt ed emissione dei neutroni di energia abbastanza
%elevata: parlare della sorgente]
We checked that the same technique, when applied to $^{252}{\rm Cf}$
runs in which prompt $\gamma$'s (selected by requiring the first
event recorded to have an energy in the range of 3 to 7 MeV) are
used to fake positrons, give results compatible with isotropic
distributions.

\section{Locating Supernov\ae}

We applied the above technique to the determination
of the neutrino direction from a Supernova
in a liquid scintillator experiment.
The difference in the neutrino energy spectrum
( the average detected neutrino energy for a Supernova
is $\sim 17\units{MeV}$ \cite{Bludman} to be compared
to the $4\units{MeV}$ of the reactor case) has two major
consequences:

%\begin{figure}[htb]
%\begin{center}
%\mbox{\epsfig{file=tnsn.ps,height=4.0in,%
%bbllx=32bp,bblly=406bp,bburx=535bp,bbury=659bp}}
%\vspace{-0.2truecm}
%\caption{Neutron angle with respect to the incident neutrino
%direction versus its kinetic energy for a Supernova neutrino
%energy distribution.}
%\label{tvsnsup}
%\end{center}
%\end{figure} 

\begin{figure}[htb]
  \begin{center}
    \mbox{\hspace{-1.0cm}\epsfig{file=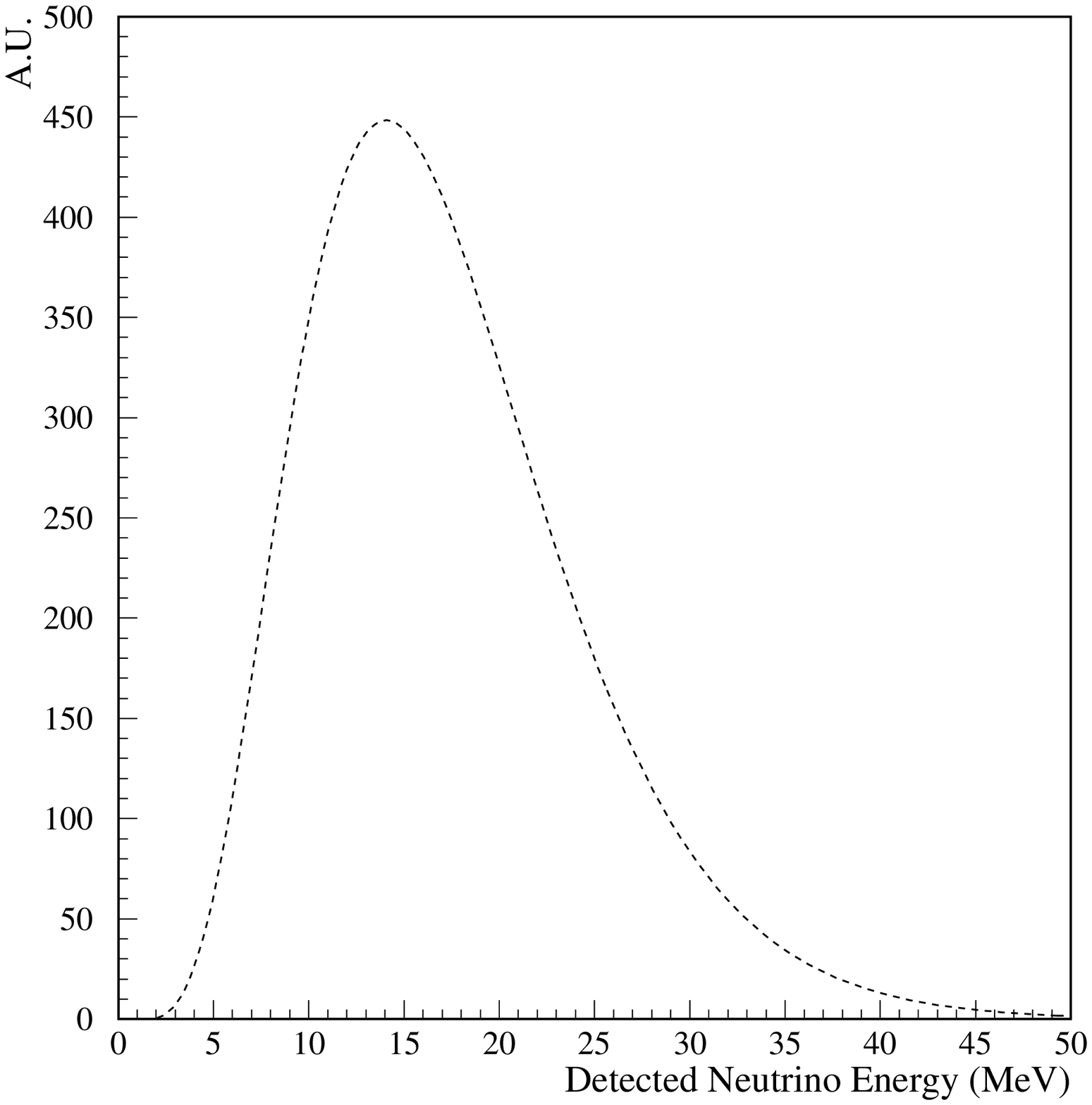,width=0.6\linewidth}
          \hspace{-0.5cm}\epsfig{file=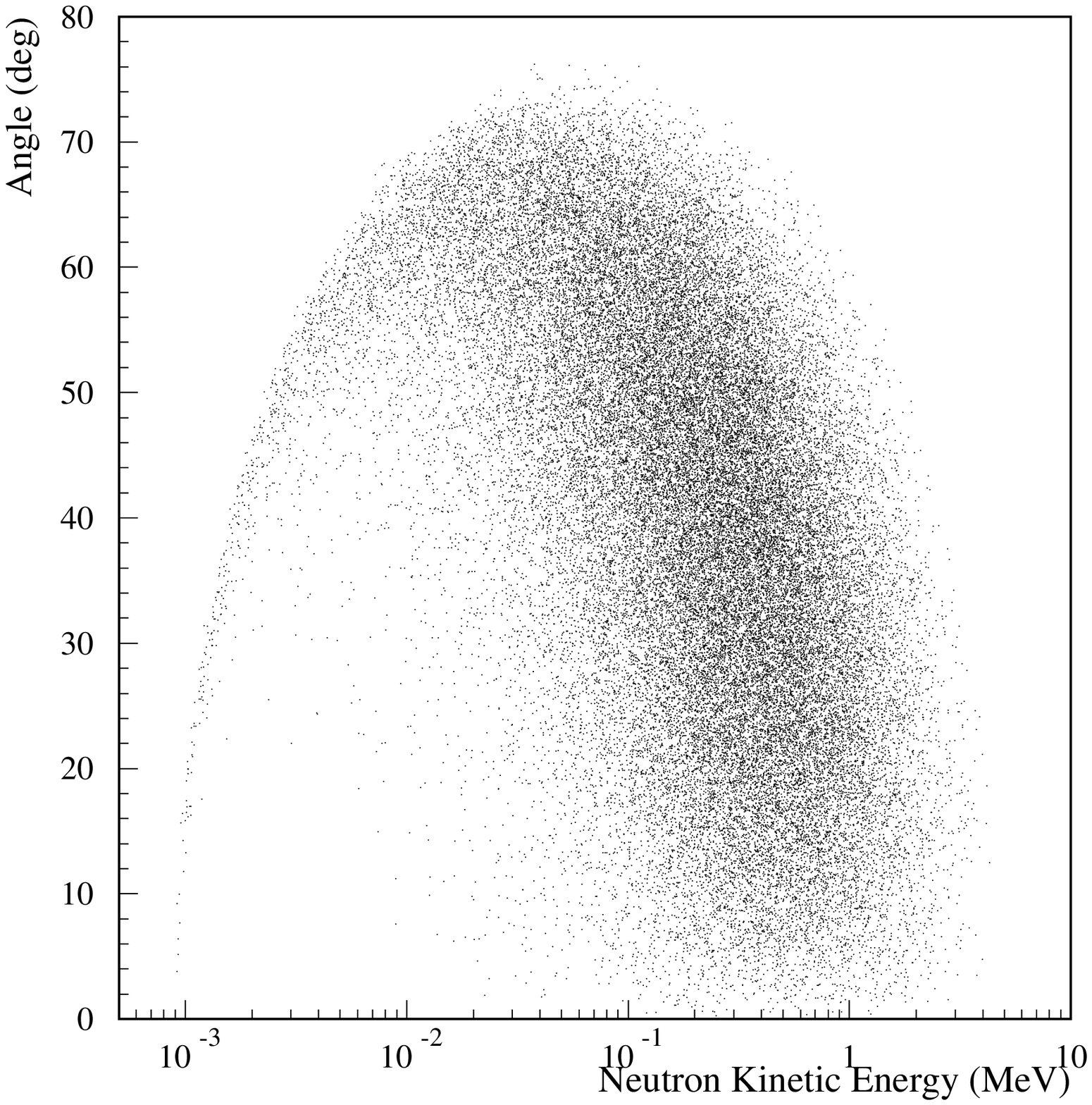,width=0.6\linewidth}}
    \caption{\small Detected Supernova neutrino energy spectrum
             (left) and neutron angle with respect to the incident
             neutrino direction versus its kinetic energy(right).}
    \label{tvsnsup}
  \end{center}
\end{figure}

\begin{itemize}
\item
the maximum neutron angle with respect to the neutrino direction
increases (in figure \ref{tvsnsup} we show the equivalent of Fig.
\ref{tvsn} for  a Supernova neutrino energy distribution)
\item
the higher neutron energy implies lower cross--sections 
which in turn imply higher displacements of the average capture
point from that of production.
\end{itemize}

Clearly these effects have opposite influences (the first, negative;
the second, positive)
on the neutrino direction determination.
The combination of the two effects was evaluated by using the 
full CHOOZ Montecarlo simulation.

We used the following Supernova $\nuebar$ 
energy distribution \cite{Bludman}:

\begin{eqnarray}
\frac{\rm dN}{\rm dE} = {\rm C}\frac{\rm E^2}{1+e^\frac{\rm E}{\rm T}}
\end{eqnarray}

\noindent 
with ${\rm T} = 3.3\units{MeV}$\footnote{Other authors\cite{Vogel1}
 use more energetic spectra which
(as we will see shortly) improve the possibility
of determininig the $\nuebar$ direction with the 
present method.}
and considered the Supernova to be at $10\units{Kpc}$.

We generated 5000 neutrino interactions in an experiment
with the same geometry, the same position resolution and the same
target (Gd-loaded liquid scintillator) as the CHOOZ experiment.
This number of events, for a Supernova at 10 Kpc, could be detected
in a liquid scintillator experiment with mass equal to that of SK.
 Such an experiment is not even foreseen
at present, but our aim here is just to compare the possibilities of
the technique we are investigating with those offered
by the neutrino elastic scattering in SK.
We chose the neutrino direction to have a  zenith angle of zero degrees
(which implies an undefined azimuthal angle).

The results of applying the technique described in paragraph \ref{technique}
to this case are shown in table \ref{ResultsMC} (first line).
The resulting
uncertainty in the direction measurement is $8.8^\circ$.
\begin{table}[htb]
    \caption{\small Measurement of the Supernova neutrino direction: 
                    Montecarlo
                    events. The results in the second line are obtained
                    by requiring the positron--neutron distance to be
                    larger than 20 cm. Note that the 
                    $\phi$ angle determination
                    is irrelevant since neutrinos are directed
                    along the zenith axis.}
    \label{ResultsMC}
  \begin{center}
    \leavevmode
    \begin{tabular}{|c|c|c|c|}

      \hline
       $\mid \vec{\rm p} \mid$ & $\phi$ & $\theta$ & Uncertainty \\
      \hline
       0.079 & $111^\circ$ & $11^\circ$ & $8.8^\circ$ \\ 
      \hline
       0.102 & $66^\circ$ & $8^\circ$ & $8.4^\circ$ \\ 
      \hline

    \end{tabular}
  \end{center}
\end{table}

From this result it follows that the increase in the 
neutron production--capture distance dominates over the broadening
of the possible neutron emission angles.
The average positron--neutron displacement in this case turns out to
be $\sim 2.5\units{cm}$.

We also investigated the effects on the direction determination
of possible event- by-event cuts on the positron energy or on the 
positron--neutron distance. The only 
statistically significant (though small) improvement 
was obtained by requiring the positron--neutron distance
to be larger than 20 cm, which corresponds to a reduction of
the neutrino sample
from $\sim 5000$ to $\sim 3400$ events.
The results obtained applying this cut are also shown in table \ref{ResultsMC}.

%We also applied our technique to the case of a 1--Kton liquid
%scintillator experiment  (by generating $\sim 300$ events) with no
%dopant element added: in this case the neutron signature is 
%the detection of the $2.2\units{MeV}\, \gamma$--ray emitted from hydrogen
%absorption. These conditions are nearer to those of the
%KAMLAND experiment. In this case the uncertainty on the
%Supernova direction turns out to be $\sim 40^\circ$, 
%dominated by the events statistics.

We finally point out that in the Supernova case the background for
Reaction \ref{reaction} is negligible since the duration
of the neutrino burst is of the order of $10\units{seconds}$.

\section{Conclusions}

The CHOOZ experiment demonstrates for the first time
the use of Reaction \ref{reaction} for measuring
the average neutrino direction.

This technique could be important in Supernov\ae\  neutrino
detection with liquid scintillator-based experiments, which
have the advantage of possessing a superior energy resolution
compared to \v Cerenkov detectors.

By using the full CHOOZ Montecarlo simulation we showed 
that in an experiment with a sufficiently high mass and
adequate position resolution,
it is possible to measure the sky coordinates
of a Supernova at 10 Kpc with
an uncertainty larger by only a factor two
($\sim 9^\circ$ compared to $5^\circ$)
than that attainable by using the angular distribution
of  neutrino elastic scattering in SK.

These results represent also an important check 
of the event reconstruction procedure of the
CHOOZ experiment and enhances our confidence in the 
analysis methods employed.

\section*{Acknowledgements}

Construction of the laboratory was funded by \'Electricit\'e de France
(EdF). Other work was supported in part by IN2P3--CNRS (France), INFN
(Italy), the United States Department of Energy, and by RFBR (Russia).
We are very grateful to the Conseil G\'en\'eral des Ardennes for having
provided us with the headquarters building for the experiment. At
various stages during construction and running of the experiment, we
benefited from the efficient work of personnel from SENA (Soci\'et\'e
Electronucl\'eaire des Ardennes) and from the EdF Chooz B nuclear plant.
Special thanks to the technical staff of our laboratories for their
excellent work in designing and building the detector.

\end{document}